\documentclass[prd,twocolumn,tightenlines,showpacs,nofootinbib,amsfonts,amssymb,amsmath]{revtex4}
\usepackage{graphicx}
\begin{document}
\newcommand{\etal}{{\it et al.}}
\newcommand{\bx}{{\bf x}}
\newcommand{\bn}{{\bf n}}
\newcommand{\bk}{{\bf k}}
\newcommand{\dd}{{\rm d}}
\newcommand{\dslash}{D\!\!\!\!/}
\def\ga{\mathrel{\raise.3ex\hbox{$>$\kern-.75em\lower1ex\hbox{$\sim$}}}}
\def\la{\mathrel{\raise.3ex\hbox{$<$\kern-.75em\lower1ex\hbox{$\sim$}}}}
\def\beq{\begin{equation}}
\def\eeq{\end{equation}}

\leftline{UMN--TH--2927/10}
\leftline{FTPI--MINN--10/33}

\vskip-2cm
\title{The Wall of Fundamental Constants}

\author{Keith A. Olive$^{1,2}$, Marco Peloso$^2$
and Jean-Philippe Uzan$^{3,4,5}$}

\affiliation{
${^1}$ William I. Fine Theoretical Physics Institute, 
University of Minnesota, Minneapolis, 55455, (USA) \\
${^2}$ School of Physics and Astronomy,
University of Minnesota, Minneapolis, 55455 (USA)\\
$^3$ Institut d'Astrophysique de Paris,
         UMR-7095 du CNRS, Universit\'e Pierre et Marie
              Curie, 98 bis bd Arago, 75014 Paris (France)\\
${^4}$ Department of Mathematics and Applied Mathematics
 Cape Town University, Rondebosch 7701 (South Africa)\\
${^5}$ National Institute for Theoretical Physics (NITheP),
Stellenbosch 7600 (South Africa). \\              
}
\vspace*{2cm}
\begin{abstract}
We consider the signatures of a domain wall produced in the spontaneous
symmetry breaking involving a dilaton-like scalar field 
coupled to electromagnetism.  Domains on either side of the wall
exhibit slight differences in their respective values of the fine-structure constant, $\alpha$.
If such a wall is present within our Hubble volume, absorption spectra at large redshifts 
may or may not provide a variation in $\alpha$ relative to the terrestrial value, depending on 
our relative position with respect to the wall. This wall could resolve the ``contradiction'' between 
claims of a variation of  $\alpha$ based on Keck/Hires data and of  the constancy of  $\alpha$ 
based on VLT data.  We derive the properties of the wall and the
parameters of the underlying microscopic model required to reproduce
the possible spatial variation of $\alpha$. We discuss the constraints on the existence
of the low-energy domain wall and describe its observational implications
concerning the variation of the fundamental constants.
\end{abstract}
 \date{November 2010}
 \maketitle
%%%%%%%%%%%%%%%%%%%%%%%%%%%%%%%%%%%%%%%%%%%%%%%%%%%%%%%%%%%%%%%%%%%%%%%%%%%%%%

There are very few observations which can be directly and 
unambiguously related to new physics. The study of relative wavelength shifts in quasar absorption 
spectra at high redshift is indeed one
of them as systematic achromatic shifts in these spectra can be attributed to changes in fundamental
constants, and in particular in the fine-structure constant, $\alpha$. This would most certainly 
call for physics beyond the standard model.
Study of the variation of constants on cosmological scales
is also the best way to test the equivalence principle on 
cosmological and astrophysical scales~\cite{jp-revue}. It
opens a window on deviations from general relativity on scales
where it is necessary to introduce dark energy and dark matter
and on which we have very little constraint on the
validity of general relativity~\cite{jp-grg}.

Claims of a variation in $\alpha$  from observations of quasar absorption spectra
using  the many multiplet method \cite{Webb} had sparked an enormous amount of 
theoretical activity in attempts to explain a temporal variation 
in the fine structure constant~\cite{massgrave,SBM,seealso,op1,cststring}. 
If confirmed, the Keck/Hires data which yielded a statistically
significant trend indicating 
$\Delta \alpha / \alpha = (-0.54 \pm 0.12) \times 10^{-5}$ over a redshift 
range $0.5 \la z \la 3.0$ (the minus sign indicates a smaller value of 
$\alpha$ in the past) could indeed point to new physics.  
However, subsequent studies based on VLT data using the same method
have shown $\Delta \alpha$ to be consistent with zero \cite{Petitjean,quast}.
These results remain somewhat controversial \cite{murphy07}.

If the low energy constants of physics depend on some
dynamical scalar field, $\phi$, they become dynamical 
and may well be space-time dependent.
On cosmological scales, it is usually thought
that the time variation dominates
over spatial fluctuations, as suggested by most models.  
The reasoning here is straightforward.
For a scalar field coupled to electromagnetism, the Lagrangian
contains a term $\frac{B_{F}(\phi)}{4}F_{\mu\nu}F^{\mu\nu}$,
$F_{\mu\nu}$ being the Faraday tensor and $B_F$ an arbitrary function
of $\phi$. This will necessarily induce a coupling to matter which is generated radiatively 
if not present at the tree level (see below).
The equation of motion for the scalar field simply takes the form
\beq
\Box \phi +
\frac{\partial V_{\rm eff}}{\partial \phi}  =0,
\label{fieldeq} 
\eeq
where $V_{\rm eff}$ includes the self interactions of $\phi$ as well as any couplings to matter,
so that it may depend on the local energy density of matter. 
For example, should the Lagrangian contain a term $B_{N}(\phi) m_N {\bar N} N$,
then the coupling to matter is effectively density dependent, which
could serve as the source of spatial variations through
\beq
\Box \phi + m_\phi^2 \phi =  B_N'(\phi) \rho_N ,
\eeq
where $m_\phi$ is the scalar mass and $\rho_N$ is the baryon energy density.~\footnote{This could lead to the mechanism now known as
chameleon~\cite{chameleon}.}
However, the density dependent shifts from the homogeneous solution
are typically extremely small except perhaps in the vicinity of 
a neutron star \cite{ellisolive,op2,barrow}. In contrast, temporal variations 
are relatively easy to achieve particularly over cosmological time scales,
as long as the field remains light.

However, there have been a series of recent puzzling observational results.
First, the combined positive Keck/Hires and negative VLT results for a change in $\alpha$
could be interpreted as a 
dipole in the spatial distribution of $\alpha$~\cite{webspace,webspace2,webspace3}.
Then, it has also been claimed that  the ratio $ m_p/m_e$ has
a small spatial variation  in the Milky Way~\cite{muspace}. 
While caution should still rule the day (the positive result for a variation in $\alpha$
has yet to be confirmed independently and may still be due to systematic effects \cite{sys}),
it is an intriguing possibility with potentially very interesting interpretations.

As we have argued above, spatial variations are expected to be much smaller than
a time variation. Indeed, if the field that triggers the variation of the constant is light during
inflation, it would have developed super-Hubble fluctuations of quantum origin, with an 
almost scale invariant power spectrum. 
The constants depending on such a field
must also fluctuate on cosmological scales and have a  non-vanishing correlation function.
This possibility is however constrained \cite{sigur}, and would not be dipole in nature.

Another possibility would be that the Copernican principle is not fully satisfied. 
Then, the background value of $\phi$ would depend e.g. on $r$ and $t$ for a spherically symmetric spacetime 
(such as a Lema\^{\i}tre-Tolman-Bondi spacetime). This could give 
rise to a dipolar modulation of the constants if the observer (us) is not located at the center of 
the universe. Such a cosmological dipole should also reflect itself on other cosmological
observations such as the cosmic microwave background (CMB) anisotropies, which does
not seem to match with the required dipole in $\alpha$~\cite{webspace,webspace2}. 
Note also that such large scale deviations from homogeneity are constrained
observationally~\cite{pctest}.

\begin{figure}[!htb]
\vskip-.4cm
\includegraphics[width=9cm]{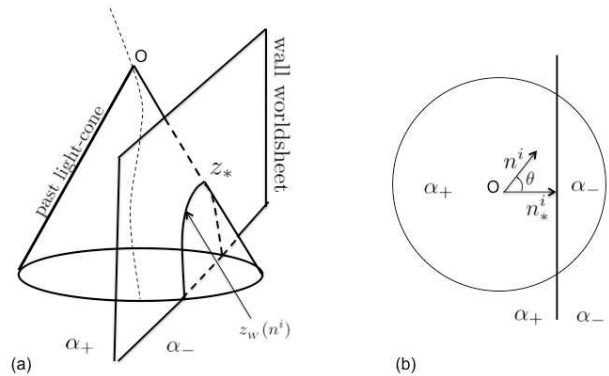}
\vskip-1.25cm
\caption{(a) A domain wall is assumed to cross our Hubble volume. It intersects
our past light-cone on a 2-dimensional spatial hypersurface characterized
by the redshift of the wall in a direction $\bn$, $z_W(\bn)$. $z_*$ is the
lowest redshift at which the wall can be observed and corresponds to a
direction $\bn_*$. According to Ref.~\cite{webspace}, $\bn_*$ should
point towards right ascension $17.3\pm0.6$~hours
and declination $-61\pm9$~deg. (b) On a constant time hypersurface, the wall cross our
Hubble horizon so that the fine structure constant takes 2 values: $\alpha_+$
in our neighborhood and $\alpha_-$ on the other side of the wall.} 
\label{fig1}
\end{figure}

Here, we propose to invoke the existence of a spatial discontinuity of the
fine structure constant, and perhaps also of other constants, due to the existence of a domain
wall crossing our Hubble volume. As depicted on Fig.~\ref{fig1} (left), if such
a domain wall exists the vacuum expectation value of the scalar $\phi$
that supports the domain wall changes sign across this hypersurface. It is then clear
that if the fine structure constant $\alpha$ is a (non-even) function of $\phi$, then it
shall take two values, one $\alpha_+$ (the larger value of $\alpha$) 
in our neighborhood and the second
$\alpha_-$ at high redshifts in the direction of the wall; see Fig.~\ref{fig1} (right). Indeed, on scales
larger than our Hubble volume, there exists a stochastic distribution of $\alpha$
taking arbitrarily one of these two values. Since in such a scenario, $\phi$
is expected to have a mass much larger than $H_0^{-1}$, it is stabilized
in one of its two vacua so that $\alpha$ has to be strictly constant in each patch
(this is similar to the landscape approach
to the cosmological constant problem but on scales of order the size
of our observable universe). This implies
that local constraints~\cite{jp-revue} on the variation of $\alpha$ such as atomic clocks,
Oklo and meteoritic dating will be trivially satisfied.

The simplest way to implement this idea is to consider
the following theory
\begin{eqnarray}
 S =\int \left[ \frac{1}{2}M_p^2 R -\frac{1}{2}(\partial_\mu\phi)^2
 +V(\phi)+\frac{1}{4}B_F(\phi) F_{\mu\nu}^2 \right. \nonumber \\ 
\left. - \sum_j  i\bar\psi_j\dslash\psi_j - B_j(\phi) m_j \bar\psi_j\psi_j
 \right]\sqrt{-g}\dd^4 x,
\end{eqnarray}
where $M_p^{-2}=8\pi G$ is the reduced Planck mass. The scalar field $\phi$ is assumed to have a simple 
quartic potential
\begin{equation}
 V(\phi) = \frac{1}{4}\lambda (\phi^2 - \eta^2)^2
 \label{quartic}
\end{equation}
and a coupling to the Faraday tensor of electromagnetism 
as well as to the fermions $\psi_j$. One could generalize the theory
so that $\phi$ couples to other gauge fields as well and even to dark matter as considered in 
Refs.~\cite{Damour,op1,SBM,cocDM}, but this would not
change our argument.
The coupling functions $B_i$ are assumed
to be of the form
\begin{equation}
 B_i(\phi) = \exp\left({\xi_i \frac{\phi}{M_*}}\right)\simeq 1 + \xi_i \frac{\phi}{M_*},
\end{equation}
where the coefficients $\xi_i$ are constant and $M_*$ is a mass scale. This model
depends on the parameters $(\lambda,M_*,\eta,\xi_F,\xi_i)$ and we shall assume here that
only $\xi_F$ is non-vanishing at tree level. 
Nevertheless, the scalar field inevitably  couples to nucleons radiatively through 
$\xi_N =  m_N^{-1} \langle N|(\xi_F/4) F_{\mu\nu}^2| N \rangle$~\cite{op1}. 
This yields $\xi_{\rm p} = -0.0007 \xi_F$ and $\xi_{\rm n} = 0.00015 \xi_F$~\cite{gl} respectively for the
proton and neutron.
Since most  baryons in the universe are protons, 
we shall take $\xi_N = \xi_{\rm p}$ for simplicity in our estimates. 

This model is a generalization of that introduced by Bekenstein~\cite{beken}
and is useful for the 
investigation of the connection between the cosmological variation of the fundamental
constants and the amplitude of the violation of free
fall in the Solar system~\cite{op1,op2,carol} and similar couplings
have been argued to be a generic prediction of string theory at low energy~\cite{cststring}. 
The main difference between the model studied here and previous models is that the scalar field
is assumed to be heavy so that it is stabilized, hence we do not
expect any local violation of the equivalence principle. Indeed, the current model
does not exhibit any temporal variation of constants once the phase transition has occurred.

The evolution of the field is dictated by the Klein-Gordon equation~(\ref{fieldeq}).
The effective potential gets three main contributions in addition to the potential~(\ref{quartic}); namely
from ({\it i}) the coupling of $\phi$ to the electromagnetic binding energy of the
matter, that is, to $\rho_{\rm baryon}$, from ({\it ii}) loop corrections that
will scale as $\xi_F^2\phi^2 T^4/M_*^2$
and from ({\it iii}) finite temperature corrections which scale as 
$(d^2 V/d\phi^2) \phi^2/24$ if the field is in equilibrium. Note that
there is no coupling to the radiation energy density since $\langle F^2\rangle=0$.
Thus, the effective potential has the form
\begin{equation}
 V_{\rm eff} = V(\phi) + \xi_N\frac{\phi}{M_*}\rho_{\rm baryon}
  + \xi_F^2\frac{\phi^2}{M_*^2} T^4 + \frac{\lambda}{8} \phi^2 T^2.
\label{V-all}
\end{equation}

To determine the typical order of magnitude of the model parameters let us first ignore the thermal 
corrections and the coupling to matter. To reproduce a change in $\alpha$ through the
domain wall matching the claimed spatial variation~\cite{webspace}, one needs
\begin{equation}
 \frac{\Delta \alpha}{\alpha} \simeq  2\xi_F  \,  \frac{\eta}{M_*}   \sim {\rm few} \, \times 10^{-6} \,.
\end{equation}
For simplicity, we shall assume in our numerical estimates that $\eta = M_*$, so 
that $\xi_F \simeq 10^{-6}$ (note that $\xi_F$ can be chosen  positive without loss of generality). 
According to the claim in Ref.~\cite{webspace}, we would need to be living in the vacuum with greater $\alpha$,
which we denoted $\alpha_+$. A greater value of  $\alpha$ means a lower value of $B_F$, which 
(for positive $\xi_F$) implies that $\phi = - \eta$ at our location. Since $\xi_p < 0$, the $\alpha_+$ vacuum has a slightly 
greater energy density than the $\alpha_-$ vacuum.

Once formed, a static domain wall has a field
configuration $\phi(z)=\eta\tanh(z/z_c)$ with $z_c= (2/\mu) = (\sqrt{\lambda/2}\,\eta)^{-1}$
being the typical thickness of the wall (this is simply the solution of the equation of motion
for $\phi$ in Minkowski spacetime and with the potential (\ref{quartic}); see e.g. 
Refs.~\cite{pubook,topdef}. It follows that its
energy density is $\rho_{\rm wall}=\lambda\eta^4/\left[2\,\cosh^4(z/z_c)\right]$ with
a surface energy density $U_{\rm wall}=2\sqrt{2\,\lambda}\eta^3/3$ obtained
by integrating $\rho_{\rm wall}$ over the transverse dimension. For a domain
wall spreading on a scale $H_0^{-1}$, the total energy density is of order
$UH_0$. It follows that the contribution of the wall  to the energy budget of the universe is of order~\cite{obswall}
\begin{equation}
 \Omega_{\rm wall} \equiv \frac{U_{\rm wall} \, H_0}{\rho_0} \simeq\left(\frac{\eta}{100\,{\rm MeV}}\right)^{3}.
\label{omega-today}
\end{equation}
where $\rho_0$ is the current total energy density of the universe and
where we have fixed $\lambda = 1$.
In the following we shall thus assume $\eta = {\cal O} \left( {\rm MeV} \right)$,  
so that the energy density in the wall is sufficiently small. It follows that the typical
values of the parameters of our model are
\begin{equation}\label{para1}
 \lambda\sim1,\quad
 \eta\sim1\,{\rm MeV},\quad
 \eta=M_*\hat\eta,\quad
 \hat\eta\sim1,
\end{equation}
and
\begin{equation} \label{para2}
 \xi_F\hat\eta\sim10^{-6},\quad
 \xi_N\sim-7\times10^{-4}\xi_F.
\end{equation}
Similarly a network of domain walls, which were however assumed to
be frustrated, with $\eta\sim100$~keV was considered in Ref.~\cite{solidDM}
as a possible explanation for the late time acceleration of the cosmic expansion.

Let us now discuss the cosmological evolution associated with the potential~(\ref{V-all}). 
In writing this potential, we have implicitly assumed that the quanta of $\phi$ have 
the same temperature as Standard Model fields. The interaction rate for the 
scattering between two photons and two quanta of $\phi$ is parametrically given by 
$\Gamma \sim\xi_F^4 \, T^5/M_*^4$. For our choice~(\ref{para1}-\ref{para2}) of parameters, 
this process is effective ($\Gamma > H$) as long as $T \ga 10 \, {\rm MeV}$. At lower temperatures 
the rate of $\phi$ self-interactions remains high, due to the  much stronger $\lambda \phi^4$ vertex.  
We therefore conclude that the potential~(\ref{V-all}) is perfectly justified.

Ignoring the small corrections proportional to $\xi_F$ and $ \xi_N$, the $Z_2$ symmetry
is restored at $T > T_{\rm C} = 2 \, {\rm \eta}$.  The third term in (\ref{V-all}) is  then 
much smaller than the fourth at any temperature of interest, and can thus be disregarded. 
The linear term in $\phi$ instead shifts the minimum to a slightly positive value of $\phi$ during the unbroken phase 
(since $\xi_{\rm p} < 0$), and introduces a tiny discrepancy between the potential of the two minima in the broken phase. At the phase transition, these effects are  negligible with respect to the relevant scale, $T$, of the $\delta \phi$ fluctuations.\footnote{For $T > T_{\rm C}$, the potential has a minimum at small but positive $\phi$. At $T<T_{\rm C}$ 
the potential has two minima, and a maximum. The local minimum and the maximum appear at $T=T_{\rm C}$, and they coincide at that time. Denoting by $\phi_-$ their common value at $T_{\rm C}$, and by $\phi_+$ the value of the true minimum at $T_{\rm C}$, we find that $\phi_+ - \phi_-  \sim 10^{-5} \, \eta \left( \xi_F \, {\rm MeV} / 10^{-6} \lambda M_* \right)^{1/3}$, and that $V \left( \phi_- \right) - V \left( \phi_+ \right) \sim 10^{-18} \, V_0 \left( \xi_F \, {\rm MeV} / 10^{-6} \lambda M_* \right)^{4/3}$, where $V_0 = \lambda \, \eta^4 / 4 \,$.} It is simple to check that the correlation length
in the standard Kibble-Zurek mechanism~\cite{topdef}, $\xi\sim T_{\rm C}^{-1}$  is much smaller than the horizon 
size at the time of the transition. The wall thus formed at a typical redshift of $1+z_{\rm f}=T_{\rm C}/T_0$
where $T_0\sim 2.348\times10^{-4}$~eV is the CMB temperature today, so that $z_{\rm f}\sim8.5\times10^{9}$.
In particular, our model differs from the late time phase transition in $\alpha$ proposed in Ref.~\cite{chacko}, where the characteristic scale in the potential is meV.

The spatial distribution of domain walls that form at the phase transition can be studied from percolation theory~\cite{perco}, which concludes that, shortly after the transition, the system must be dominated by a large wall with an extremely complicated structure spreading along the entire Universe \cite{walldyn}. Smaller closed walls are also present, but they quickly contract and decay. Typically, the evolution of the system eventually leads to one large wall per Hubble radius. In the case considered here,  the two minima on different sides of the wall have different energy due to the linear term in $\phi$ present in (\ref{V-all}).  As a consequence, the wall is subject to a force towards the region of higher potential. In our case, this corresponds to the wall moving towards our location. 

Due to this, a number of consequences may be expected if the wall moves at a relativistic speed today. Firstly, the absorption regions, from which the variation of $\alpha$ is deduced, need to be in the $\alpha_-$ vacuum, while we need to be in the $\alpha_+$ vacuum. Light passing through these regions and arriving to us needs to cross the wall. This is highly constrained if the wall itself is moving towards us at relativistic speed. Secondly, photons crossing the wall can be reflected by it with an ${\cal O} \left( \delta \alpha / \alpha \right)^2$ probability (this can be easily obtained by computing the reflection and transmission coefficients of a flux of photons incident on the wall); a further $\left( \xi_N / \xi_F \right)^2$ suppression is present for matter. Even if it is a small probability, the reflected photon (or nucleon) will have a large energy if the $\gamma$ factor of the wall is high. This could lead to phenomenological signatures when, for example, the wall crosses a star. Thirdly, the angle of the cone in which the cone is visible, Eq.~(\ref{cone}) below,  is affected if the wall moves relativistically. Note, however, that
interestingly the term in $B_F(\phi)F^2$ does not affect the equation of
propagation of photons in the eikonal approximation at
first order~\cite{sur}.

\begin{figure}[!htb]
\includegraphics[width=6cm,angle=-90]{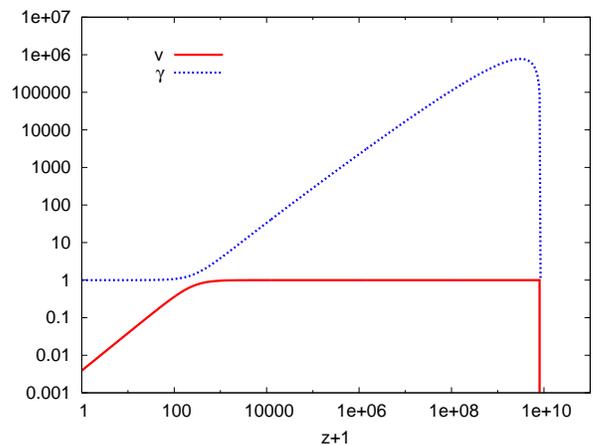}
\caption{Evolution of the physical velocity of the wall $v$ (solid line) and of the associated boost factor $\gamma$
(dotted line), plotted as a function of cosmological redshift $z$. We recall that time evolves from right to left in the Figure.} 
\label{fig2}
\end{figure}

The motion of the wall can be derived from 
the dynamics of extended objects~\cite{wallforce} according to $T_{\rm wall}^{\mu\nu}K_{\mu\nu}=f$
where $T_{\rm wall}^{\mu\nu}$ and $K_{\mu\nu}$ are
the stress-energy tensor and extrinsic curvature of the wall and $f$ the force acting on it. This
reduces to
\begin{equation}
\frac{d^2 \, x^3}{d \tau^2} + \Gamma_{\mu \nu}^3 \, \frac{d x^\mu}{d \tau} \, \frac{d x^\nu}{d \tau}
= \frac{\gamma}{R} \,  \frac{\Delta V}{U_{\rm wall}}
\label{eq-motion}
\end{equation}
for the motion of a flat wall in the transverse direction $x^3 \equiv z$,
where $\tau$ is the proper time measured by an observer on the wall, $R$ the scale factor of the universe, $\Delta V = 2 \eta \xi_N \rho_{\rm baryon} / M_*$ is the potential difference between the two sides 
(see Ref.~\cite{bias} when the effects of wall curvature are relevant). 
As $x^3$ is a comoving coordinate, the physical velocity of the wall is given by
$v = R \, {d x^3}/{d t} = ({R}/{\gamma}) \, {d x^3}/{d \tau}$
where $t$ is the physical time, and  $\gamma \equiv d t / d \tau = 1/\sqrt{1-v^2}$. In terms of physical
quantities, Eq.~(\ref{eq-motion}) can be rewritten as
\begin{equation}
\frac{d }{ d t} \left(R\gamma \, v \right) = R \frac{\Delta V}{U_{\rm wall}}\equiv R \,  F.
\end{equation}
From the expressions for $\Delta V$ and $U_{\rm wall}$, and from the current baryon density
$\rho_{{\rm baryon},0} \simeq 1.8 \times 10^{-48} \, {\rm GeV}^4$, we deduce that
$F=2.7 \times 10^{-48} \, \left( 1 + z \right)^3\, f \, {\rm GeV}$ with
\begin{equation}\label{eq-evol-wall}
f \equiv \left(\frac{\xi_F}{10^{-6}}\right) \left(\frac{\eta}{1~{\rm MeV}}\right)^{-3}\hat\eta\,\lambda^{-1/2}.
\end{equation}
This equation of motion can be solved numerically assuming that the wall starts at rest at $T=T_{\rm C}= 2 \, {\rm MeV}$,\footnote{We can show analytically that the velocity of the wall today depends only  logarithmically on the initial temperature.} and for $f=1$; the result is depicted in Fig. \ref{fig2}. We see that, right after its formation, the wall accelerates to a large boost factor, $\gamma \sim {\cal O} \left( 10^{6} \right)$. However, it gradually slows down due to Hubble friction. The current peculiar velocity of the wall is $v_0 \simeq 0.004$, so that none of the effects mentioned above is an issue.
We also see that, in the non-relativistic regime, the velocity of the wall scales linearly with the parameter $f$.

To be viable, our model must satisfy an additional set of constraints that we now summarize.
\begin{enumerate}
\item{\it CMB constraints}. While the effects of cosmic strings on the
CMB have been extensively studied~\cite{CMBstring}, domain walls have 
not been widely considered since they were thought to be formed at much higher energy and thus
have a dramatic effect on the CMB. 

A single wall would contribute to the temperature anisotropy via the integrated Sachs Wolfe effect 
\cite{Zeldovich:1974uw}. For a static universe, and in the non-relativistic regime, the gravitational potential generated by a very large (planar) wall grows linearly with the distance $L$  from the wall
(notice that this has a zero net effect if the distance between the source and the wall is equal to that between the wall and the observer). For a typical cosmological distance $ \sim H_0^{-1}$, we estimate $\delta T / T \sim G \, U_{\rm wall} \,  H_0^{-1}$. This results in 
\begin{equation}
 \left(\frac{\delta T}{T}\right)_{\rm CMB}\sim 10^{-6}\left(\frac{\eta}{1\,{\rm MeV}} \right)^3.
\end{equation}
which constrains $\eta$ to be smaller than a few MeV. Another source of temperature anisotropy is related to the fact that $\alpha$ is not constant across the visible universe. This is not an issue for the model we are discussing, since the cmb data are only sensitive to a variation $\Delta \alpha / \alpha  \sim {\cal O} \left( 10^{-2} \right)$ or greater \cite{cmb}. Finally, we can also neglect the 
temperature anisotropy related to the probability that a CMB photon is not transmitted through the wall, 
which is of $ {\rm O } \left(\Delta \alpha / \alpha \right)^{2}$, as we have already mentioned.

\item{\it Astrophysical constraints.}  Although our scalar is relatively heavy, $m_\phi \sim 1$ MeV,
it can be produced in supernovae.  
The production rate of scalars through inverse decay is roughly,
\beq
\Gamma_{\gamma\gamma\to \phi} \sim \frac{\xi_F^2}{M_*^2}\,T^3 .
\label{sigma}
\eeq
In principle for scalars with mass $m_\phi < T$, this could result in an excessive energy loss rate.
However, these scalars decay to two photons 
with a rate $\Gamma_d \sim \xi_F^2 \mu^3/M_*^2 \sim \xi_F^2 M_*$.
Requiring that their decay length is smaller than the size of the core leads to a lower 
bound on their mass
\beq
\frac{M_*}{1\,{\rm MeV}} > \mathcal{O}(10^{-2}) \times \left( \frac{10^{-6}}{\xi_F} \right)
\eeq
(for typical energies of order $T \sim 30 $ MeV).
Thus, for our choice~(\ref{para1}-\ref{para2}) of parameters, these scalars decay within the core and there is no energy loss.  

\item{\it Tunnelling to the true vacuum}. Contrary to a standard domain wall, the non-minimal
coupling induces a shift between the two minima.  The lifetime
of the false vacuum is of the order~\cite{coleman} 
$$
 \tau\sim \Lambda\exp\left(\frac{27\pi^2}{8}\frac{S_0^4}{\Delta V^3} \right)
$$
with $S_0=\int_{-\eta}^{\eta}\sqrt{V(\phi)}\dd\phi\sim 2\eta\sqrt{\Delta V}$.
Now, with $\Delta V\sim \xi_N\Omega_{m0}\rho_{\rm crit}(\eta/M_*)(1+z)^3$ 
and assuming $\Lambda\sim M_*^{-1}$, we conclude that
$\tau=H_0^{-1} f_\tau$ with
$$
f_\tau= \frac{H_0}{M_*} \exp\left(\frac{54\pi^2}{\Omega_{m0}\xi_N}
\frac{M^4_*}{\rho_{\rm crit}}\hat\eta^3(1+z)^{-3}
\right).
$$
The argument in the exponential is always very large and scales
as $4.5\times10^{54}\hat\eta^3(1+z)^{-3}$. Even at the time the wall is formed, 
$z_{\rm f}\sim 10^{10}$,  the factor is larger than
$10^{24}$.
This means that the wall forms at a time where the false vacuum has
a lifetime larger than our Hubble time. Since the lifetime of the wall increases with
time, we are guaranteed that today, the wall is effectively stable.
\end{enumerate}

\begin{figure}[htb!]
\vskip .2in
\includegraphics[width=8cm]{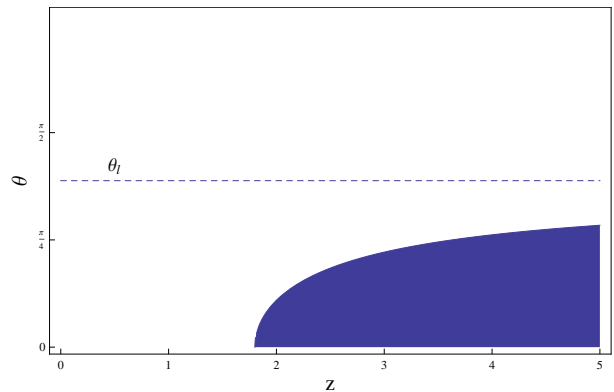}
\caption{Summary of the observable prediction of our model. Assuming the wall
is at a redshift $z_*\sim1.8$ at its closest position to us in a direction $\bn_*$, then
the fine structure constant is a function of $\cos\theta=\bn_*\cdot \bn$ and $z$. The
blue region corresponds to $\alpha=\alpha_-$ and the upper
region to $\alpha=\alpha_+$. The fact that our observable universe has a finite
radius implies that we shall detect no variation for angle larger
than $\theta_l$.} \label{fig3}
\vskip -.2in
\end{figure}

Our model also has some specific observational predictions that arise from the fact
that $\alpha$ takes two discrete values. Let us denote $\bn_*$ the direction of the wall
and $z_*$ the redshift of its closest position and $\chi_*=\chi(z_*)$ the comoving
radial distance to which it corresponds.  Since the equation of our
past-light cone is $\chi=\eta_0-\eta$ and that of the wall $\chi=\chi_*/\cos\theta$ if
it is assumed to be at rest in the cosmological rest frame. It follows that the redshift
of the wall in a direction $\bn$ is given by $\chi(z)=\chi_*/\cos\theta$
with $\cos\theta= \bn_* \cdot\bn$. Solving this equation gives the dependence of $\alpha$
with $z$ and $\theta$ and is plotted in Fig.~\ref{fig3}. Moreover since our observable
universe as a finite radius, the discontinuity can be observed only in a cone
of angle $\theta_l$ around the direction $\bn_*$ with
\begin{equation}\label{cone}
 \cos\theta_l =\chi_*/\chi_{\rm H}.
\end{equation}
$\theta_l$ depends on $z_*$ and on the cosmological parameters. Typically,
$\cos\theta_l\sim0.345$ if $z_*\sim1.8$.\\

In summary, we have proposed a two vacuum solution to produce a large scale spatial
variation of the fine structure constant, which is not
associated to a local time variation. In the model presented, the scalar field
was coupled to electromagnetism and to nucleons through the photon 
contribution to the nucleon mass.  It is of course quite plausible that 
this coupling extends to other gauge fields (and perhaps Yukawa couplings) as well.
In that case, we should expect not only spatial variations in $\alpha$, 
but also coupled spatial variations in other quantities such as 
particle masses and $\Lambda_{\rm QCD}$ \cite{ellisolive,co}. 

For example, one might expect a variation in the proton-to-electron mass ratio, $\mu$.
If due to coupled variations of fundamental constants, one typically expects
that it is correlated to the variation of $\alpha$; $\Delta\mu/\mu
\sim -50 \Delta \alpha/\alpha$~\cite{opq}.  Molecular hydrogen transitions in quasar absorption 
systems yield a limit $\Delta\mu/\mu = (-2.6  \pm 3.0) \times 10^{-6}$ \cite{king}. Searches for
spatial variations in the proton to electron mass ratio in the Milky Way \cite{muspace} produced
a non-zero result of $\Delta\mu/\mu = (2.2 \pm 0.4 \pm 0.3) \times 10^{-8}$
and conservatively implies an upper limit of $3 \times 10^{-8}$. Analogous searches for
a spatial variation of $\alpha^2\mu$ yield an upper limit of $3.7 \times 10^{-7}$
\cite{Fspace}.  Taken together, these would imply an upper limit of $\Delta \alpha/\alpha < 2 \times 10^{-7}$.  In the context of the model presented here, this limit would present no 
difficulty with the Keck/Hires observations indicating a variation in $\alpha$.
The Milky Way being entirely in the $\alpha_+$ vacuum would show no variations in any of these
quantities and these observations therefore could not constrain variations in
the $\alpha_-$ vacuum. 

We also note that we might expect a small effect on the light element abundances
produced in big bang nucleosynthesis \cite{coc}. Whether or not the transition 
happened before or after BBN, the average light element abundance
should correspond to abundances using constants determined
at $\phi = 0$ (or at a slightly smaller value of $\alpha$).  However variations at the level of 
$10^{-6}$ would hardly be perceptible in element abundances. 
Similarly a small change in $\alpha$ would very slightly affect 
the CMB through recombination~\cite{cmb}, but this too would be imperceptible. 
The same may be true for future constraints that can be set from
signals originating from the absorption of the CMB at 21 cm hyperfine transition of the neutral atomic hydrogen~\cite{21cm} or from stellar evolution~\cite{ekstrom} with sensitivities which are
typically of the order of $10^{-2}$ for $z>30$ and $10^{-5}$ for
$z\sim15-20$ respectively.

Other standard powerful systems~\cite{jp-revue} for testing the variation in $\alpha$ are the Oklo 
phenomenon and meteoritic  studies. But as these are both on our side of the wall, we would expect 
that they are too local to probe any variation in $\alpha$.  Similarly, we would not expect any positive result
from atomic clock measurements of changes in $\alpha$ unless of course the 
wall passes through the solar system in the near future.  Given the
sensitivity of these experiments (see e.g., Ref.~\cite{rosenband}), there would be no
mistaking the event.
 
\vskip.25cm
\noindent{\bf Acknowledgements:} 
We would like to thank C.R. Contaldi, G.F.R. Ellis, J. Murugan, P. Peter, 
C. Ringeval, and M. Voloshin 
for helpful discussions.
The work of KAO and MP was supported in part by
DOE grant DE-FG02-94ER-40823 at the University of Minnesota.
The work of JPU was supported by a PEPS-PTI grant from CNRS (2009-2011)
and the PNCG (2010).

\end{document}